# Large-scale signatures of unconsciousness are consistent with a departure from critical dynamics

Enzo Tagliazucchi[1,2*], Dante R. Chialvo[3*], Michael Siniatchkin[1], Enrico Amico[4], Jean-Francois Brichant[4], Vincent Bonhomme[4], Quentin Noirhomme[4], Helmut Laufs[2,5*], Steven Laureys[4*]

* These authors contributed equally to this work

[1] Institute for Medical Psychology, Christian Albrechts University Kiel, 24105 Kiel, Germany.

[2] Department of Neurology and Brain Imaging Center, Goethe University Frankfurt am Main. Frankfurt am Main, 60528 Frankfurt am Main, Germany.

[3] Comision Nacional de Investigaciones Cientificas y Tecnologicas (CONICET), Argentina.

[4] Coma Science Group, GIGA Research and Cyclotron Research Center, University and University Hospital of Liège, Liège, Belgium.

[5] Department of Neurology, Christian Albrechts University Kiel, 24104 Kiel, Germany.

**Contact information:**
Enzo Tagliazucchi
Institute for Medical Psychology, UKSH
Arnold-Heller-Straße 3
24105 Kiel
Germany
Tel: 0157 - 81976950

tagliazucchi.enzo@googlemail.com

**ABSTRACT**

Loss of cortical integration and changes in the dynamics of electrophysiological brain signals characterize the transition from wakefulness towards unconsciousness. In this study we arrive at a basic model explaining these observations based on the theory of phase transitions in complex systems. We studied the link between spatial and temporal correlations of large-scale brain activity recorded with fMRI during wakefulness, propofol-induced sedation and loss of consciousness, and during the subsequent recovery. We observed that during unconsciousness activity in fronto-thalamic regions exhibited a reduction of long-range temporal correlations and a departure of functional connectivity from anatomical constraints. A model of a system exhibiting a phase transition reproduced our findings, as well as the diminished sensitivity of the cortex to external perturbations during unconsciousness. This framework unifies different observations about brain activity during unconsciousness and predicts that the principles we identified are universal and independent from its causes.

# 1. INTRODUCTION

Anesthetic drugs transiently impair awareness and thus offer a unique opportunity to investigate the neural correlates of conscious wakefulness. In contrast to other reversible unconscious states (such as sleep), anesthetics simultaneously reduce arousal and awareness and – except in the rare event of intra-operative awareness – result in a brain state incompatible with conscious content (Alkire et al. 2008). Studies of the transition from wakefulness to loss of consciousness induced by propofol (a presumed GABA agonist anesthetic agent) consistently report decreased cortical integration (Alkire et al. 2008; Lee et al. 2009; Boveroux et al. 2010; Schrouff et al. 2011; Monti et al. 2013; Amico et al., 2014) and changes in the dynamics of electrophysiological brain signals, such as delta (1 - 4 Hz) and gamma oscillations (30 - 70 Hz) (Murphy et al. 2011; Boly et al. 2012). Despite many experimental reports at different temporal and spatial scales, the precise mechanisms underlying propofol-induced unconsciousness remain poorly understood. The need for a mechanistic understanding of this phenomenon is non-trivial, since it could contribute to unraveling how consciousness is constructed and preserved by the brain.

During conscious wakefulness the cortex spontaneously generates a flurry of ever changing activity (Chialvo 2010; Raichle 2011; Sporns 2011). In the temporal domain, this activity is characterized by long-range temporal correlations, meaning that signal fluctuations at the present time influence dynamics up to several minutes in the future (Maxim et al. 2005; He 2011). Lacking any distinctive scale (scale-free), these temporal correlations can be characterized by the computation of scaling exponents, such as the Hurst exponent (Tagliazucchi et al. 2013a). In the spatial domain, these fluctuations are coordinated across networks of regions commonly co-activated during stimulation and cognitive performance, termed resting state networks (RSN) (Beckmann et al. 2005; Smith et al. 2009). While functional connectivity can transiently dissociate from inter-areal anatomical connections (Liegeois et al. 2015), brain activity correlations computed over extended periods of time seems to be partially explained by the underlying anatomy (Hagmann et al. 2008; Hermundstad et al. 2013; Greicius et al. 2009; Wang et al. 2013). This suggests that spontaneous brain activity can be understood as an ever-transient (or metastable) exploration of the wide repertoire of paths offered by the underlying structural connectivity, the extent of such exploration potentially depending on the brain state, with variable repertoires corresponding to different degrees of awareness.

Here we put forward an interpretation of propofol-induced loss of consciousness in analogy to the dynamics and connectivity of fluctuations seen on a diversity of complex systems exhibiting different phases. As a system explores the space of possible configurations, its spatio-temporal correlations behave in characteristic ways. In particular, the dynamical changes underlying different degrees of awareness could be analogous to the qualitative changes observed in the dynamics of complex systems when they move away from a phase transition (Werner 2013). Experimental evidence gathered from functional magnetic resonance imaging (fMRI) data supports the view that during conscious wakefulness the human brain operates near the critical point of such a transition (Chialvo 2010; Tagliazucchi et al., 2012). A robust feature of the critical state is the phenomenon of *critical slowing down*, which is manifest as increased temporal autocorrelation (i.e. long-range temporal correlations) of fluctuations throughout the system (Werner 2007; Chialvo 2010; Kelso 2012). Far from the critical transition, the variables describing the system are very stable. As a consequence, any perturbation from equilibrium is dissipated quickly (i.e. dynamics rapidly returns to equilibrium). Conversely, near the transition the effects of any disturbance last longer, thus it is that the dynamics *slow down*. Since far from the transition the system is stuck into a stable state, its dynamics cannot explore the wider repertoire allowed by structural constraints. The opposite occurs near the phase transition at which the system can switch between a large number of locally stable or metastable states (Werner 2007), and fully explore its structural connectivity (Stam et al., 2014; Deco et al., 2014). Thus, if unconsciousness results in a departure from critical dynamics, we expect to see these two inter-related signatures: 1) Loss of temporal correlations in brain activity time series and 2) a less complete exploration of the activity patterns allowed by the underlying structural connectivity.

Previous experimental results are consistent with a loss of critical slowing down in large-scale brain activity during unconsciousness. For instance, magnetic and electric perturbation of the cortex during different states of consciousness elicits equally different responses: conscious wakefulness is characterized by prolonged and spatiotemporally correlated responses (disturbances last longer), whereas unconsciousness is characterized by a smaller repertoire of rapidly vanishing and spatially localized responses (Massimini et al. 2005; Ferrarelli et al. 2010; Casali et al. 2013; Pigorini et al. 2014). The response to endogenous fluctuations during deep sleep is also rapidly vanishing, resulting in the loss of temporal long-range correlations (Tagliazucchi et al. 2013a). Finally, spontaneous electrophysiological activity recorded during unconsciousness presents increased stability (Solovey et al. 2015). A mechanistic account of the action of propofol on large-scale brain activity should provide a unified explanation for these seemingly different experimental results.

To propose such an explanation, we studied fMRI data acquired during wakefulness, propofol-induced sedation and loss of consciousness, as well as during the subsequent recovery of awareness. We evaluated the presence of two large-scale signatures of a departure from the critical point of a phase transition: loss of long-range temporal correlations and the uncoupling of functional and anatomical connectivity (Stam et al., 2014; Deco et al., 2014), measured using diffusion tensor imaging (DTI) and diffusion spectrum imaging (DSI).

Finally, we developed a conceptual model presenting a phase transition to assist in the mechanistic interpretation of the experimental results.

## 2. METHODS

*Experimental design and participants*

Participants were scanned with fMRI during wakefulness (W), propofol sedation (S), propofol-induced loss of consciousness (LOC) and finally during the recovery of wakefulness (R). Sedation corresponded to Ramsay level 3 (Ramsay et al. 1974). Loss of consciousness corresponded to Ramsay levels 5-6 (subjects did not exhibit responses to verbal instructions). Recovery corresponded to Ramsay level 2.

Twenty healthy right-handed volunteers aged between 18 and 31 years (22.4± 2.4 years) were initially included in the study. Following Monti et al (2013), subjects with head displacements exceeding 3 mm during any of the four conditions were discarded from the analysis, resulting in a final set of 12 participants. For all conditions, the resulting average head movement amplitudes did not exceed 1 mm (wakefulness: 0.38 mm, sedation: 0.25 mm, loss of consciousness: 0.17, recovery: 0.36 mm). No significant effect of condition on head displacement was found ($F_{3,44}$= 2.63, p=0.062). As noted by Monti et al., this is a conservative approach to limit the impact of head movement. Other methods, such as scan nulling (Power et al. 2014), could affect the estimation of blood-oxygen level dependent (BOLD) signal spectral power and long-range temporal correlations and therefore were not applied. As an additional control, the presence of significant residual correlations between absolute and relative head movement time series and voxel-wise BOLD time series was evaluated after data pre-processing, with no significant residual correlations being detected.

Details on fMRI, DTI and DSI data acquisition and preprocessing are provided in the Supplementary Methods.

*Estimation of long-range temporal dependencies*

DFA (Kantelhardt et al. 2001) was applied to study the temporal correlations of BOLD fluctuations. This method was developed to obtain estimates of long-range temporal dependence in time series, while accounting for the possibility of non-stationarities. In the Supplementary Methods we provide a formal definition of the procedure followed in the DFA algorithm. Briefly, time series were first de-trended by subtracting the mean and the cumulative sum was then computed. Afterwards, the signal was divided into non-overlapping windows and the intensity of the fluctuations was computed by averaging the standard deviation of the signal across all windows (de-trended within each window). This procedure was repeated for different window sizes and the slope of the standard deviation of the fluctuations vs. the window size ("fluctuation function", in logarithmic scale) was identified with the Hurst exponent (H). Based on the value of H, three qualitatively different scenarios can be distinguished: long-range temporal correlations (slow decay of the autocorrelation function) with 0.5 < H < 1, uncorrelated temporal activity (exponential decay of the autocorrelation function) with H = 0.5 and long-range anti-correlations (switching between high and low values in consecutive time steps) with 0 < H < 0.5.

We applied DFA to the first 150 volumes of the BOLD time series of every voxel for each subject and condition, obtaining spatial maps of H values. To compute H, windows of length 10, 15, 25 and 30 volumes were used, as the logarithmic plot of the fluctuation function showed linear behavior within this range. We also estimated H in the frequency domain following a wavelet-based method. The steps followed for the wavelet estimation of H are extensively presented and discussed in the Supplementary Methods.

*Functional network construction*

We constructed functional networks by extracting average BOLD signals from all regions of interest and computing the linear correlation between all pairs of signals, resulting in the correlation matrix $C_{ij}$.

For comparison with the underlying anatomical connectivity networks, the correlation matrices $C_{ij}$ were thresholded to yield binary adjacency matrices $A_{ij}$ such that $A_{ij} = 1$ if $C_{ij} \geq \rho$ and $A_{ij} = 0$ otherwise. The parameter ρ was chosen to fix the ratio of the connections in the network ($\sum_{i>j} A_{ij}$) to the total possible number of connections (termed link density). It is important to fix the link density when comparing networks since otherwise differences could arise because the means of the respective $C_{ij}$ are different (and therefore the number of non-zeros entries in $A_{ij}$) and not because connections are topologically re-organized across conditions.

We performed all analyses for a range of link densities between 0.01 and 0.3 in steps of 0.01. When comparing functional networks with their anatomical counterparts, the chosen link density ranges always included the link density of the DTI and DSI anatomical networks.

*Similarity between functional and anatomical connectivity neighborhoods*

We defined the connectivity neighborhood of node i as $n_j = A_{ij}$ (i.e. the i-th column of the adjacency matrix for a fixed local link density). According to this definition, the j-th entry of $n_j$ is 1 if nodes i and j share a direct connection in the network, and it is zero otherwise. We obtained the connectivity neighborhood of all nodes in the anatomical and functional networks across all conditions and participants, as well as for a range of local link densities. To estimate the similarity between the anatomical and functional connectivity neighborhoods of each node we computed the Hamming distance between the anatomical and functional versions of vectors $n_j$ (normalized by their total length). The Hamming distance is defined as the number of symbol substitutions (in this case 0 or 1) needed to transform one sequence into another and vice-versa, and in this case it is equal to twice the number of connections that must be re-wired to turn the functional connectivity neighbor into the anatomical connectivity neighbor.

*Fluctuations in functional connectivity and repertoire of functional networks*

We investigated if the fluctuations in the transient connectivity within the frontal executive control RSN were more widespread during wakefulness vs. propofol-induced unconsciousness by computing the average functional connectivity of all nodes in the RSN over short non-overlapping windows of different durations. Afterwards, we computed the variance of the time series of dynamical functional connectivity fluctuations.

We investigated the repertoire of functional networks explored over time by means of a new methodology (see Fig. S7 of the Supplemental Information for a schematic). We first computed the connectivity matrices of all nodes within the executive control RSN (Fig. S7A) over non-overlapping segments of 20 volumes. After thresholding at a given link density (ranging from 0.01 to 0.4) this defined a series of binary networks explored over time (Fig. S7B). Afterwards, we computed the average correlation between the adjacency matrices of all these binary networks (Fig. S7C). If the repertoire of explored networks is very constrained this average correlation is high (i.e. all transient networks are very similar). On the other hand, if the system explores a wide range of different transient networks, this average correlation is lower. We termed this index transient network similarity (TNS) index.

*Computational model*

The computational model is based on the previous work of Haimovici et al. 2013 (see also a posteriori similar formulation by Stam et al. 2014). It consists of an underlying anatomical network of connections (DSI network) and rules for the transition between three states: inactive, active and refractory. The rules for the transitions at the i-th node are as follows,

1) Inactive to active: either spontaneously with a probability of $10^{-3}$ or if $\sum_{j \text{ is active}} W_{ij} > T$.
2) Active to refractory always occurs.
3) Refractory to inactive with a probability of $10^{-1}$.

These rules were used to simulate time series that were subsequently binarized by setting the active state to 1 and the other two to 0, and convolved with the standard hemodynamic response function mimicking the brain neurometabolic coupling. As shown in Haimovici et al. 2013, a second order phase transition exists at $T_C \approx 0.05$. At this point, activity becomes self-sustained, spatial and temporal correlations are maximized and an optimal agreement with the empirical fMRI data is obtained (including an approximate reproduction of the major RSN reported in the work of Beckmann et al. 2005).

## 3. RESULTS

We first obtained Hurst exponent and the low-frequency (0.01 - 0.1 Hz) power for each participant and condition (W, S, LOC and R). Also, we investigated the same metrics in a phantom made of water (see Figure S1 of the Supplementary Figures).

The anatomical distribution of Hurst values and low frequency power reflected the division of cortical anatomy into grey and white matter and cerebrospinal fluid. BOLD signals from grey matter voxels were characterized by long-range temporal correlations (H > 0.65) whereas white matter and cerebrospinal fluid voxels generally presented relatively weaker temporal correlations and 0.01-0.1 Hz frequency power (Fig. 1A). A shift towards reduced H and low-frequency power can be observed in the LOC condition (third row). We computed the global Hurst exponent and low-frequency power values (averaged across all grey matter voxels) and observed reduced values for the LOC condition relative to W (Fig. 1B). We also observed reduced values of the metrics in the frequency domain for R relative to W, suggesting that the recovery from propofol-induced loss of consciousness might not have been complete. Histograms for H and low-frequency power are shown in Fig. S1 (Supplementary Figures). H values peaked at around 0.5 (corresponding to temporally uncorrelated dynamics) for the water phantom and at H > 0.5 for grey matter brain voxels, i.e. as opposed to brain dynamics, those of the water phantom were temporally uncorrelated.

We conducted voxel-wise statistical tests to assess the effect of the condition on H and low-frequency power (Fig. 2). We observed a significant effect of the condition (W, S, LOC and R) on H (both DFA and wavelet-estimated) and 0.01-0.1 Hz power. This was observed in in a set of regions comprising the thalamus, the ventromedial and orbitofrontal cortices, the frontal and rolandic operculi, the superior and medial frontal gyri and the anterior cingulate and bilateral insular cortices. Post-hoc t-tests between W and all other conditions revealed significant decreases only for the comparison vs. LOC. Similar results can also be observed in the first-order autoregressive coefficient of BOLD signals (Fig. S2, Supplementary Figures). Statistical parametric maps are presented in Fig. 2A (bottom panel). Fig. 2B shows a ranking of the top ten automated anatomical labeling (AAL) atlas (Tzourio-Mazoyer et al. 2002) regions based on the statistical significance of the contrast W vs. LOC. The extent of the overlap between the three different metrics is shown in Fig. 2C as a joint rendering of differences in H (both DFA and wavelet-estimated) and 0.01-0.1 Hz power. No significant differences were observed in terms of the goodness of fit ($R^2$) of the DFA fluctuation function. The covariance between the statistical significance maps derived from all three metrics is shown in Fig. S3 (Supplementary Figures).

We then studied the coupling between anatomical and functional connectivity. At first, we restricted both functional and anatomical connectivity networks to a sub-network encompassing the executive control network reported in Beckmann et al. 2005, since this RSN overlapped with the regions where we found a breakdown of long-range temporal correlations during LOC (see Fig. S4 of the Supplementary Figures). For both DTI and DSI anatomical connectivity networks and almost all link densities we observed decreased similarity between anatomical and functional connectivity networks during LOC relative to W (Fig. 3A).

Afterwards, we studied the local similarity between the anatomical and functional first neighbors of all individual nodes in whole-brain networks. The network nodes associated with decreased anatomical-functional coupling during LOC relative to W are shown in Fig. 3B. Differences encompassed the thalamus, as well as the medial prefrontal cortex, anterior cingulate cortex, frontal and rolandic operculi and the bilateral insular cortex. A ranking of AAL regions by their percentage of nodes with significant differences is presented in Fig. 3C. The robustness of the results with respect to the two anatomical connectivity networks is manifest in the joint rendering of the nodes presenting significant differences (Fig. 3D). Similar results were obtained using partial correlations instead of linear correlations (see Fig. S6 of the Supplementary Figures).

As discussed in the introduction, we hypothesized that during LOC the de-correlation of temporal dynamics should be seen together with a less thorough exploration of the repertoire of possible states allowed by anatomical constraints. To address this possibility, we investigated whether changes in H and low-frequency power during LOC were correlated with the degree of anatomy-function coupling. We computed the average anatomy-function Hamming distance within the significant regions in Fig. 3B (bottom panel) as a function of the link density, as well as the average H (DFA and wavelet-estimated) and 0.01-0.1 Hz frequency power in the same regions. This was performed for each participant in the LOC condition. We then computed the correlation coefficients and associated p-values between H, low frequency power and the mean Hamming distance as a function of the link density. Results are shown in Fig. 4A (note that this correlation is against structural-functional network distance, not similarity). For both anatomical connectivity networks and almost all link densities, a significant negative correlation between H and the mean Hamming distance was found. Correlations involving low frequency power were also negative but in most cases slightly above the threshold of statistical significance. Negative correlations imply that the stronger the de-correlation in temporal dynamics, the stronger the uncoupling between anatomical and functional connectivity. Fig. 4B shows example scatterplots obtained at the reference link density of 0.15.

We then investigated the variability of functional connectivity over time to determine if unconsciousness was characterized by diminished fluctuations in dynamic connectivity, as predicted by a departure from criticality (see Haimovici et al., 2013). Results presented in Fig. 5A reveal that the variance of functional connectivity fluctuations (over a wide range of window sizes) were diminished during propofol-induced loss of consciousness. Furthermore, a wider range (repertoire) of functional networks was explored during conscious wakefulness compared to unconsciousness (Fig. 5B), as quantified by the TNS index computed using windows of 20 volumes.

To further gauge the significance of our observations we introduced a simple dynamical model to evaluate which qualitative aspects of the propagation of information in anatomical networks were more relevant to replicate our empirical observations. The model allows three possible states for each node in the DSI network. The possible node states and transitions between them are illustrated in Fig. S5 of the Supplementary Figures.

The threshold in the model controls the propensity of excitations to propagate throughout the anatomical network. Values higher than the critical threshold of $T_C \approx 0.05$ difficult the propagation of activity, which eventually dies out. On the other hand, lower thresholds result in self-sustained activity. Very low values result in the extreme of many nodes becoming rapidly activated and then transitioning towards the refractory ("hyperpolarized") state. A critical point exists at $T_C \approx 0.05$, marked by self-sustained activity allowing the reproduction of many features of large-scale brain activity, such as long-range temporal correlations in space and time and the emergence of coordinated structures strongly resembling RSN (Haimovici et al. 2014). The critical point corresponds to a second order phase transition, characterized by maximal variability in the intrinsic dynamics of the system, critical slowing down and an optimal exploration of the repertoire of metastable state (i.e. states in which the system transiently resides). Examples of the temporal dynamics during the sub-, super-, and critical regimes are shown in Fig. S5 of the Supplementary Figures.

We found that the similarity between functional and structural connectivity was maximal near the critical point. This was evident from computing the correlation between functional and anatomical adjacency matrices at each threshold value (Fig. 6A, left) or by computing the Hamming distance between the binarized functional and anatomical connectivities of each node (Fig. 6A, right). The frequency at which activations occurred correlated negatively with the threshold. As shown in Fig. 6B (left), low values facilitated the propagation of activity and induced higher activation rates while higher thresholds caused the opposite effect by hindering the propagation of activity. In the super-critical ($T > T_C$) regime, the frequency of activations also correlated negatively with the anatomical-functional distance (Fig. 6B, right). The same result was observed for the Hurst exponent of the average activity generated by the model. Both are consistent with the changes observed under propofol: the higher the uncoupling between anatomical and functional connectivity, the faster and less temporally correlated the dynamics of the system.

The critical slowing down observed when dynamics are close to the phase transition maximizes the response of the model to external perturbations. We studied the average response of the system to a sudden excitation of 60% of the nodes. This computation is of interest to evaluate which phases of the model better correspond to the diminished response to magnetic perturbations of cortical activity observed during loss of consciousness. In Fig. 7 (left) we show the average time course after a perturbation (computed over 100 simulations) both for $T_C \approx 0.05$ and for $T_C < T = 0.01$. The response in the critical case was characterized by a sustained oscillation, with temporally persistent activity observed after the perturbation. On the other hand, the perturbation in the super-critical case induced a transient response rapidly giving way to a baseline of uncorrelated oscillations. We measured the decay of the variance in the activity over short temporal windows of 20 time steps. The activity decay after the perturbations is shown in Fig. 7 (center) for all thresholds. By measuring the time elapsed until a level of low variance ($10^{-5}$) was crossed, we estimated the time necessary for the activity to decay to its baseline. The decay time peaked near the critical point and quickly decreased both in the super- and sub- critical cases (Fig. 7, right).

## 4. DISCUSSION

We studied how propofol-induced loss of consciousness affected the temporal dynamics of BOLD signals and how the changes in large-scale dynamics were related to the exploration of the underlying anatomical connectivity. Loss of consciousness was paralleled by a shift towards faster and temporally uncorrelated BOLD signals in the frontal lobe, the salience network and the thalamus. Within the same regions, functional connectivity departed from the underlying anatomical constraints; this departure covaried with loss of long-range temporal correlations.

An interpretation for our results is given in Fig. 8. We show a schematic depiction of an elementary system composed of interacting units, the state of the system being symbolized by the position of a particle within a potential landscape with several local equilibria (potential wells). In reality, this potential would span a high-dimensional space, with the state vector describing a multitude of independent variables characterizing the system at each time point.

However, we adopt this simplified schematic for illustration purposes. Far from the critical point of a phase transition (left panel), the system is more stable and the local minima are deeper; in consequence any external perturbation or internal fluctuation rapidly vanishes and the particle returns quickly to the same local equilibrium. For the same reason, the dynamics of the system do not allow the exploration of all possibilities offered by the structural connectivity and thus functional correlations reflect only a portion of the anatomical connections. Near the phase transition (right panel) the landscape becomes shallower, the stability decreases and perturbations can induce a more widespread exploration of the potential landscape (see also Fig. 5), resulting in more sustained changes. As the system explores the neighborhood of different local equilibria (or metastable states) spatial correlations better reproduce its structural connectivity. Our observations of large-scale fMRI dynamics and connectivity during loss of consciousness can be interpreted as a departure from a critical state (near the transition) towards more stable fluctuations (far from the transition).

Our model also allowed us to simulate the effect of perturbations near and far from its phase transition and thus to connect two robust but seemingly unrelated findings characterizing states of reduced awareness: loss of temporal complexity (i.e. long-range temporal correlations [Tagliazucchi et al. 2013a]) and rapidly vanishing responses to direct magnetic and electric stimulation of the cortex (Massimini et al. 2005; Ferrarelli et al. 2010; Casali et al. 2013; Pigorini et al. 2015). Within our framework, both arise as a result of increased stability, with endogenous as well as exogenous fluctuations failing to displace the system between different metastable states.

The mechanisms by which propofol could result in dynamics compatible with a departure from a phase transition deserve further investigation. Most likely, these consist of alterations in the properties of individual units (neurons or groups of neurons) translating into dramatically different collective behaviors. For instance, our model suggests that facilitated spreading of activity results in a state of global hyperpolarization (see Fig. S5) impairing the propagation of external perturbations throughout the system. A possible correlate of this facilitated spreading is the increased power in the gamma frequency band observed during propofol-induced unconsciousness (Murphy et al. 2011; Boly et al. 2012), which is also a main driver of BOLD activity fluctuations (Nir et al. 2007).

Contemporary theories postulate that consciousness is an emergent phenomenon of physical processes in the brain. The explanation of subjective experiences from the objective observation of these processes has remained elusive to neuroscience. However, it is possible to ask what features of brain activity are compatible with the rich subjective phenomenology of consciousness. An aspect common to different theories is that consciousness can be associated to a state of high *neural* complexity (Tononi et al., 1994; Tononi and Edelman, 1998). This can be understood as a state between the extremes of very high differentiation without information integration (the dynamics of each unit in the system become independent, like in a "disordered" or "random" system at the super-critical state) and very low differentiation (the system presents few possible states, as in an "ordered" or "regular" sub-critical system). At the - between ordered and disordered - critical state, dynamics are both integrated (the units of the system present long-range correlations both in time and space) and segregated (the system allows the exploration of a large number of possible metastable states), suggesting this is the state that could maximize the standard definition of neural complexity. Future work will need to formally address a possible equivalence between metrics of *neural complexity* and metrics of criticality (i.e. order parameters).

Our research provides evidence that the "baseline" state of wakeful rest presents critical dynamics and that unconscious brain states depart from this kind of dynamics. Thus, we identify critical dynamics with the state of consciousness. Since the possibility of having conscious, reportable content ("*I see X, hear Y, feel Z*") is in general conditional to being in a conscious state; critical dynamics could also be a necessary requirement for conscious content to emerge. This is supported by the observation of comparable neural complexity at rest and during conscious information access (Burguess et al., 2003). Furthermore, the role critical dynamics play in conscious information access could be related to the observation that at criticality the response of the system to external stimuli is maximized (see Fig. 7). Since conscious perception requires the engagement of a distributed set of neurons (*dynamical core*; Tononi and Edelman, 1998, as well as in the concept of the *global workspace*; Dehaene and Naccache, 2001), a prerequisite is a high sensitivity to incoming stimuli (high *susceptibility*). Conversely, at the sub- or super-critical states, sensory stimulation results in a local and transient perturbation failing to propagate to more widespread networks related to conscious perception.

At other spatial and temporal scales, evidence for an association between consciousness and critical dynamics has been obtained in the context of deep sleep (Priesemann et al. 2013), anesthesia (Alonso et al. 2014; Scott et al. 2014) and epileptic seizures (Meisel et al. 2012).
Here we introduced the coupling between anatomical and functional connectivity as a signature of the critical state, which is particularly fit for fMRI recordings since both can be measured at the same spatial resolution using this technique. Propofol-induced loss of consciousness resulted in diminished anatomical-functional coupling, interpreted here as a departure from the critical regime characteristic of conscious wakeful rest (Tagliazucchi et al 2012, Expert et al. 2010). Changes in function-anatomy uncoupling can be understood in terms of the emergence of long-range correlations at criticality. The term "long-range" must be treated with caution when discussing the human brain, since regions far away in Euclidean space may be close together in a topological sense (i.e. directly connected anatomically). Thus, we did not expect to see a breakdown of long-range functional connectivity as a function of Euclidean distance following the departure from criticality, but a separation from the structural connectivity backbone instead.

Recent work on anesthetized primates (Barttfeld et al. 2015A) found that transient patterns of functional connectivity strongly resembling the anatomical constraints were more frequent during loss of consciousness relative to wakefulness, which appears to contradict our results. However, two important differences must be taken into account. First, we found diminished anatomy-function coupling over *extended* periods of time, which is related to the average of the functional connectivity states visited over time (as an analogy, mapping the average route traced by cars in a city throughout a entire day, as opposed to taking instantaneous snapshots). Second, the effect we report was regionally localized to a set of frontal regions and the thalamus, as opposed to the global effect reported in Barttfeld et al. 2015A. This last distinction is very important since the richness of anatomical connectivity varies throughout the brain, from the regular structure of the cerebellum and primary cortices to the highly complex, variable and phylogenetically advanced frontal and parietal associative cortices (Kaas et al. 2013; Rilling et al. 2014). Indeed, we observed that the functional exploration of fronto-thalamic anatomical connectivity was hindered under propofol, highlighting its importance for the maintenance of conscious awareness. We note that decreased similarity between anatomical and functional connectivity could also result from the selective enhancement of functional connections that are not associated with structural links (as an analogy, cars taking "shortcuts" across regions not directly connected by roads). This possibility is ruled out by the breakdown of within- and between-network functional connectivity observed during propofol-induced unconsciousness (Boveroux et al. 2010).

We also studied the dynamics of BOLD signals, which have received comparatively less attention in the context of anesthesia than electrophysiological recordings. We observed a departure from slow and temporally correlated dynamics in frontal regions and in the thalamus. These areas strongly overlap with those where decreased metabolism under anesthesia was reported (Alkire et al. 1997; Kaisti et al. 2002; Laitio et al. 2007; Bonhomme et al. 2008). Breakdown of long-range temporal correlations was also reported in other unconscious brain states such as deep non-rapid eye movement (NREM) sleep (Tagliazucchi et al. 2013a). This led us to hypothesize that long-range temporal correlations of spontaneous activity fluctuations are a primary characteristic of brain activity during conscious wakeful rest. Phenomenologically, the subjective feeling of continuity during conscious wakefulness ("stream of consciousness", as famously phrased by William James [James, 1980]) cannot be supported by short-range temporal correlations as exhibited, for example, in Markovian dynamics (when the state of the system depends only on the immediately previous state). The short-term persistence of conscious information is impossible under these dynamics unless structural changes occur, which likely belong to a completely different temporal scale (Bullmore and Sporns 2009).

The main limitation of our manuscript arises from the indirect nature of fMRI recordings and the possibility of propofol influencing other physiological variables that are not directly related to the level of consciousness. Experimental evidence shows that the effects of propofol on arterial blood pressure and cerebral blood flow are small (Fiset et al., 2005; Liu et al., 2013; Veselis et al., 2005; Johnston et al., 2003), ruling out confounds related to pressure-dependent changes in BOLD signals. As discussed by Hudetz and colleagues (Hudetz et al., 2015), confounds due to alterations in neurovascular coupling are also unlikely given the preservation of functional responses during propofol-induced loss of consciousness (Franceschini et al., 2010). Experiments measuring cardiac and respiration rates simultaneously with fMRI during propofol-induced loss of consciousness did not find a significant difference vs. conscious wakefulness (e.g. Schröter et al., 2012). Another possibility is that our results reflect the concentration of propofol in blood but not the responsiveness of the participants (Barttfeld et al., 2015B). One argument against our results reflecting the increasing concentration of propofol in blood is the fact that we did not observe any significant effects under propofol-induced sedation (a state characterized by responsiveness in spite of non-zero propofol plasma concentration). Our results were specific to unconsciousness, as determined by the onset of state of unresponsiveness (Ramsay et al., 1974).

In summary, we achieved an empirical characterization of large-scale brain activity during propofol-induced unconsciousness in terms of inter-related changes in spatial and temporal correlations. In analogy to other complex systems undergoing phase transitions, the dynamics became temporally uncorrelated during unconsciousness and failed to efficiently explore the underlying structural connections. Since the proposed interpretation is based on general principles of complex systems, further research should reveal the universality of our findings across other brain states of diminished awareness, as well as investigate their applicability for the objective assessment of levels of consciousness.

## ACKNOWLEDGMENTS

This work was funded by the Bundesministerium für Bildung und Forschung (grant 01 EV 0703) and the LOEWE Neuronale Koordination Forschungsschwerpunkt Frankfurt (NeFF). We thank Ben Palanca and two anonymous reviewers for valuable comments on this manuscript, Ed Bullmore and Nicolas Crossley for sharing the DTI data and Patric Hagmann and Olaf Sporns for sharing the DSI data.

**FIGURE CAPTIONS**

**Figure 1: Anatomical specificity of long-range temporal correlations and low frequency (0.01-0.1 Hz) fluctuations.** (A) Anatomical overlays of the mean Hurst exponent (DFA and wavelet estimation) and low frequency power for all experimental conditions; long-range temporal correlations and low frequency fluctuations were predominantly observed in cortical and sub-cortical grey matter. (B) Differences in global Hurst exponents and low frequency power relative to the values measured during wakefulness (*$p<0.05$, Bonferroni corrected for multiple comparisons).

**Figure 2: Breakdown of long-range temporal correlations and reduced low frequency power fluctuations during propofol-induced loss of consciousness.** (A) Top: Main effect of experimental condition (wakefulness, sedation, loss of consciousness and recovery) on the Hurst exponent (DFA) and low frequency power. Bottom: Reduced Hurst exponent and low frequency power during loss of consciousness compared to wakefulness. Both statistical significance maps were thresholded at $p<0.05$, FDR-controlled for multiple comparisons. (B) Regions in the AAL atlas ranked according to their differences between wakefulness and loss of consciousness. (C) Combined anatomical overlay of the three metrics presented in Panel A.

**Figure 3: Regional dissociation of anatomical and functional connectivity during loss of consciousness.** Results in the left column were obtained using the DTI network with 401 nodes, those in the right column using the DSI network with 998 nodes. (A) Similarity (correlation coefficient) between anatomical and functional connectivity networks within the executive control RSN as a function of link density, obtained during wakefulness (blue) and loss of consciousness (red). (B) Anatomical overlay of regions with significant increases in anatomical-functional distance during loss of consciousness vs. wakefulness. (C) Ranking of AAL regions according to the percentage of nodes they contained with significant differences in anatomical-functional distance. (D) Joint anatomical rendering of results obtained using the DTI and DSI anatomical connectivity networks.

**Figure 4: Changes in long-range temporal correlations and anatomical-functional coupling are correlated during loss of consciousness.** (A) Left: Correlation coefficient between the Hurst exponent (DFA and wavelet estimation) and low frequency power averaged over the regions in Fig. 3B, and the average distance between anatomical (DTI) and functional connectivity; results are shown as a function of the link density. Right: same computation for the DSI network. (B) Scatter plots of anatomy-function distance vs. H (DFA and wavelet estimation) for a reference link density of 0.15 (light blue dashed line in panel A).

**Figure 5: Unconsciousness increases network stability and decreases the repertoire of transient network states.** (A) The variance of transient average connectivity within the executive control RSN as a function of non-overlapping window size, for wakefulness and loss of consciousness. (B) TNS index (computed using non-overlapping windows of 20 volumes) for wakefulness and loss of consciousness.

**Figure 6: Dynamics and connectivity of the model.** (A) Left: Similarity between anatomical and functional (i.e. simulated) connectivity as a function of the threshold T. Right: Hamming distance between anatomical and functional node connectivity neighborhoods (averaged across all nodes) as a function of T. In both cases, the highest anatomical-functional coupling is observed close to the critical point (T = $T_C$). (B) Left: Frequency of node activations as a function of T. Right: Mean anatomical-functional distance as a function of frequency of node activations and Hurst exponent of average activity of the model. As in the experimental data, higher frequency of activations and diminished long-range temporal correlations paralleled the dissociation between anatomical and functional connectivity patterns.

**Figure 7: The sensitivity to external perturbations is maximal near the phase transition of the model.** (A) Average time course after a perturbation (activation of 60% of the nodes) during the critical and supercritical regimes. (B) Decay of activity after a perturbation for a range of thresholds. (C) Decay time as a function of T. The longest decay times are obtained when T = $T_C$.

**Figure 8: Schematic representation of the dynamics of a system far and near the critical point of a phase transition**. The state of the system at a given time is represented by the position of the particle on the potential landscape U(x). The system at equilibrium (green) is perturbed at $t_0$ subsequently relaxing (red, $t_1 \rightarrow t_2$) at different speeds depending on whether it is far (left) or near (right) a phase transition. Far from the transition (left) the system is stable and the local minima (equilibrium points) are deep, consequently dynamics are rapidly restored and the effects of perturbation are short-lasting. Near the phase transition (right) the landscape of the potential is shallow and consequently the stability of the local minima decreases, which is reflected in the time domain (middle panels) as a slowing down of the system response to fluctuations. The change in stability can be also observed spatially because it affects the exploration of the different metastable states of the system. The graphs denoted by SC (structural connectivity) represent a portion of the underlying structural network. The bottom diagram (FC; functional connectivity) denotes the structural paths traversed during an interval of time. For shallow local minima the structural paths leading from node A to the rest of the nodes (B, C, D, E) are equally likely, resulting in a complete exploration of the underlying structural connectivity. On the contrary, far from the transition (left) the potential barriers separating the nodes are taller, leading to the observation of only a portion (in this case only one) of all possible paths.

**Figure 1**

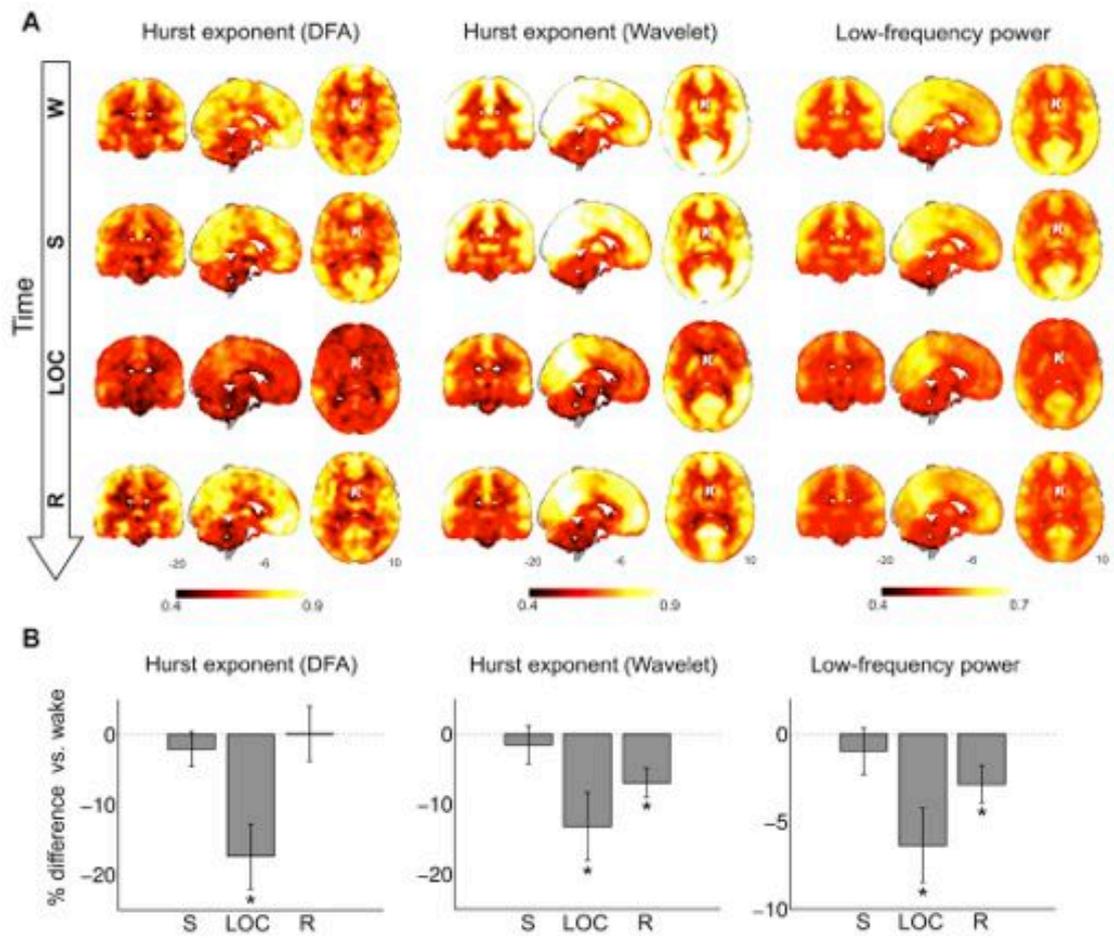

**Figure 2**

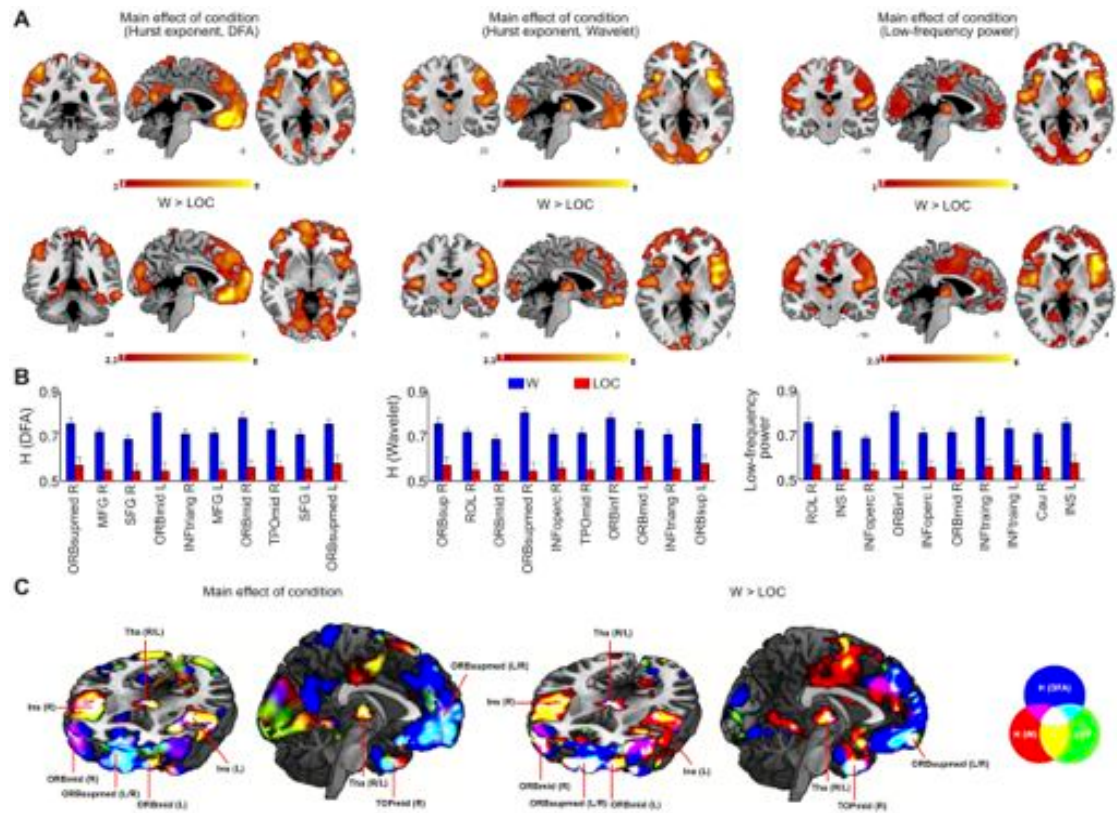

**Figure 3**

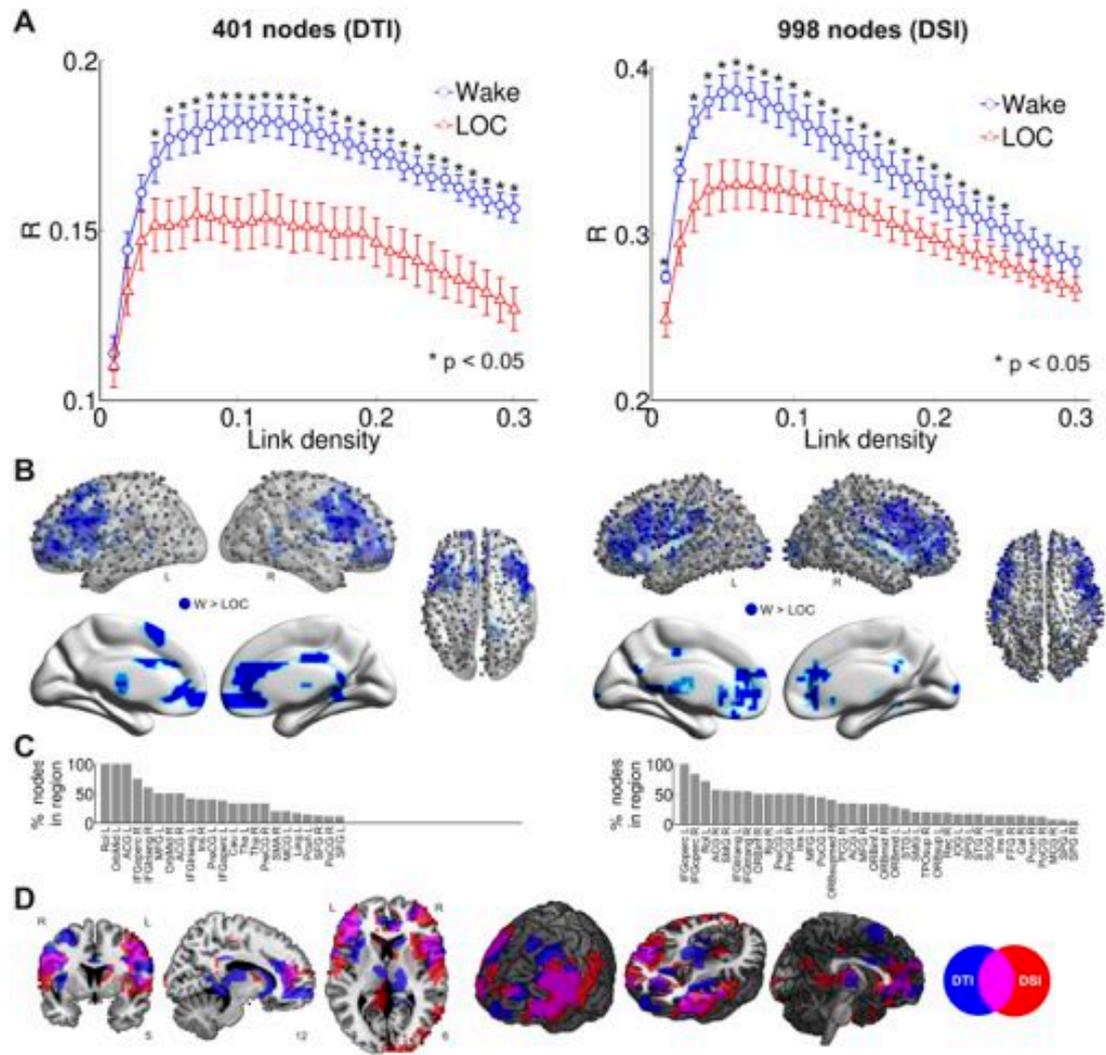

**Figure 4**

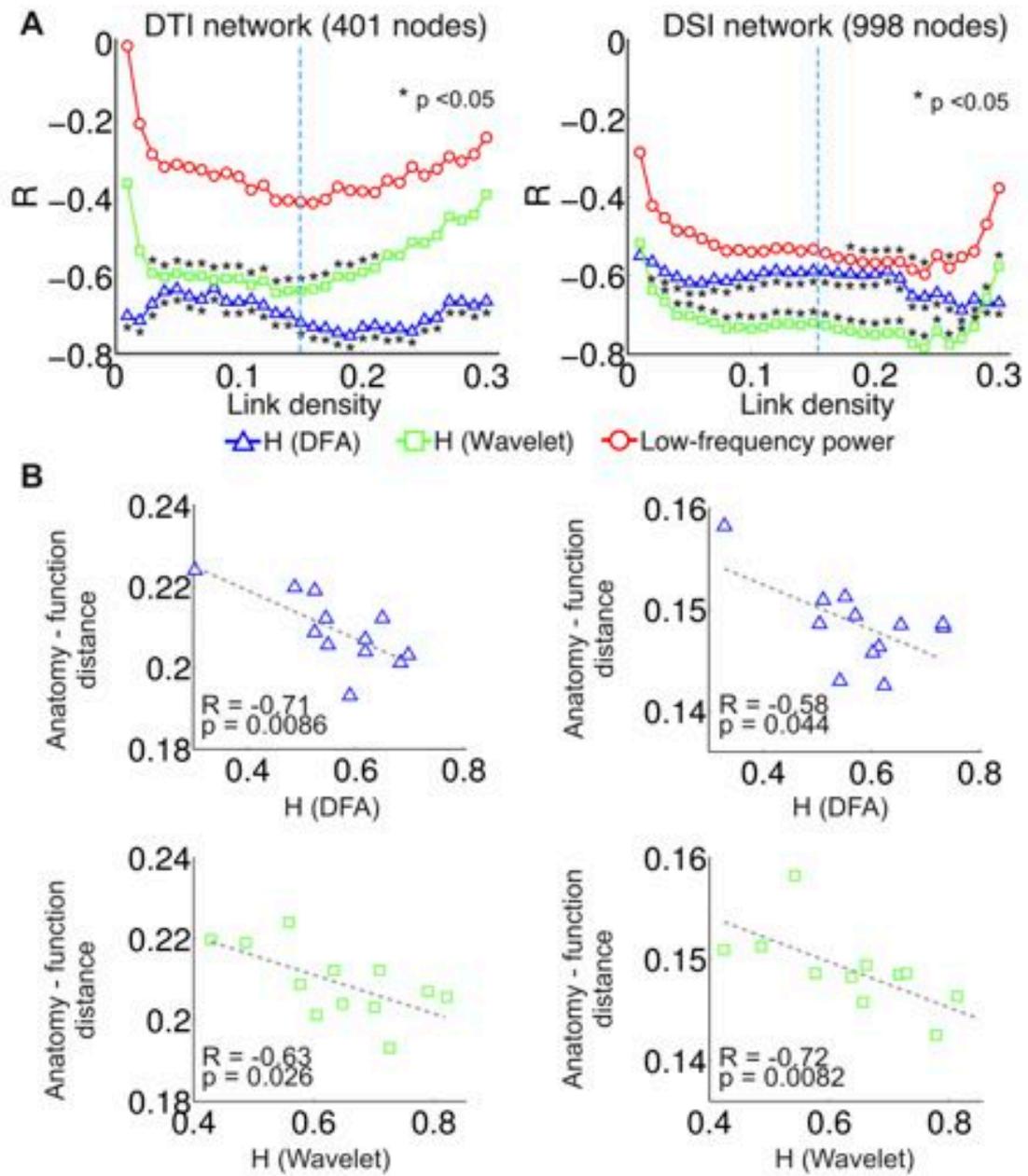

**Figure 5**

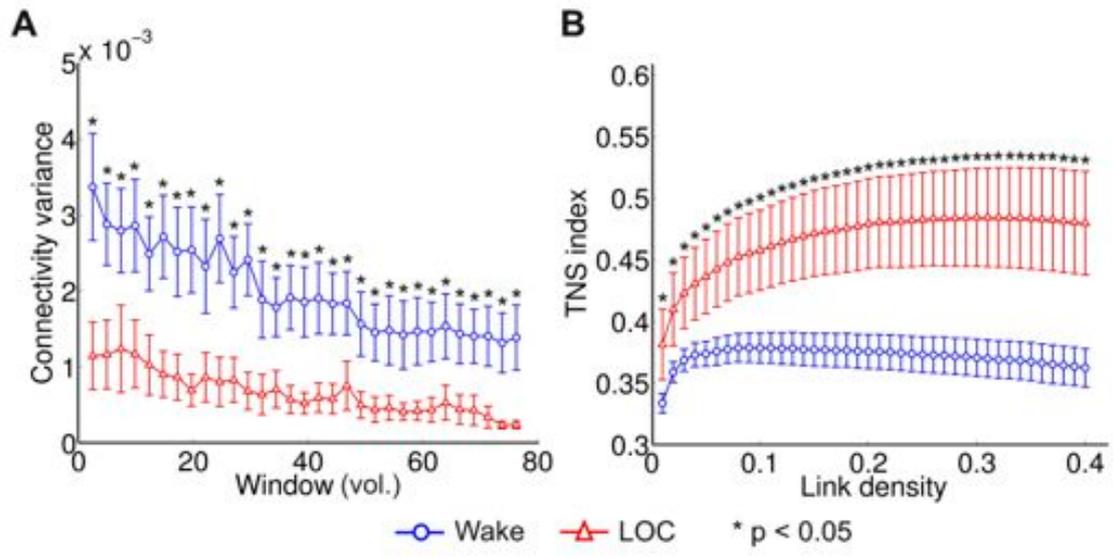

**Figure 6**

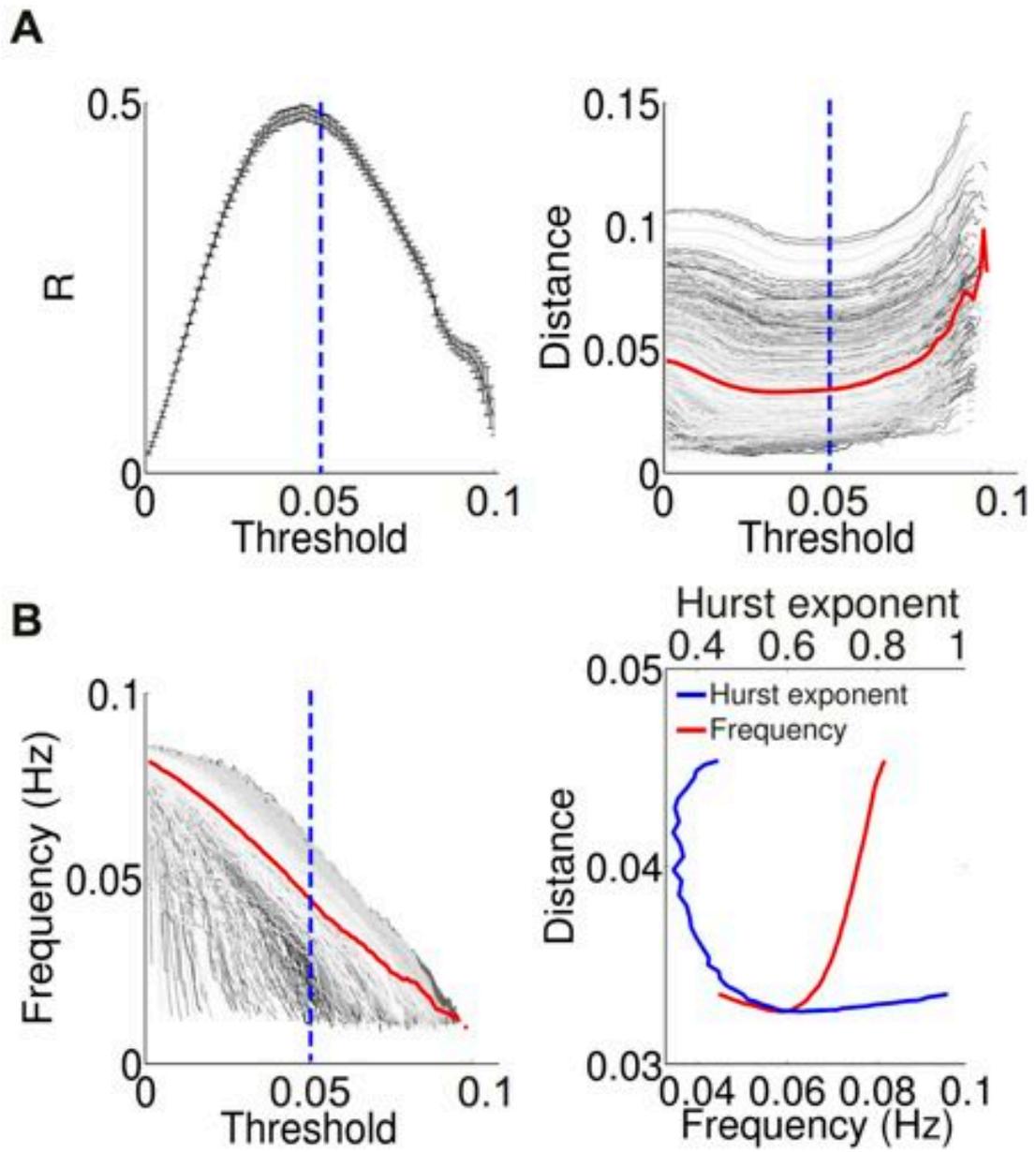

**Figure 7**

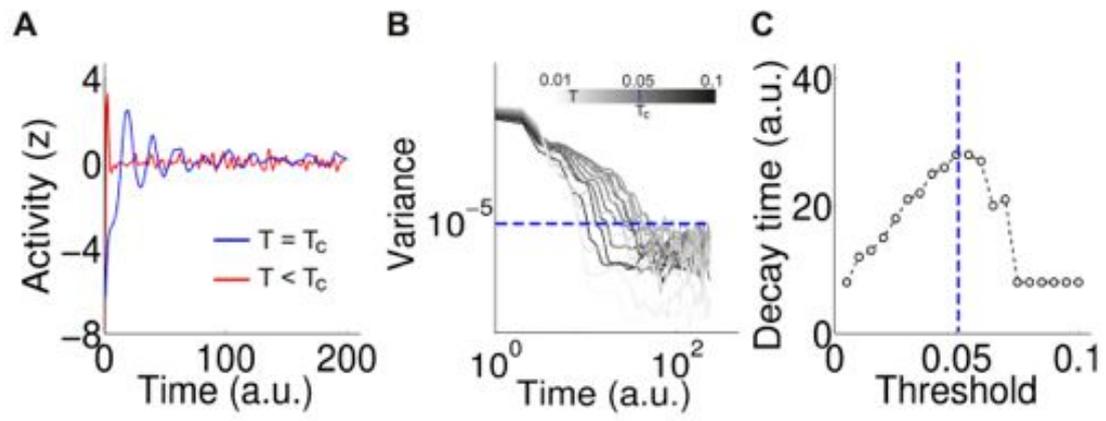

**Figure 8**

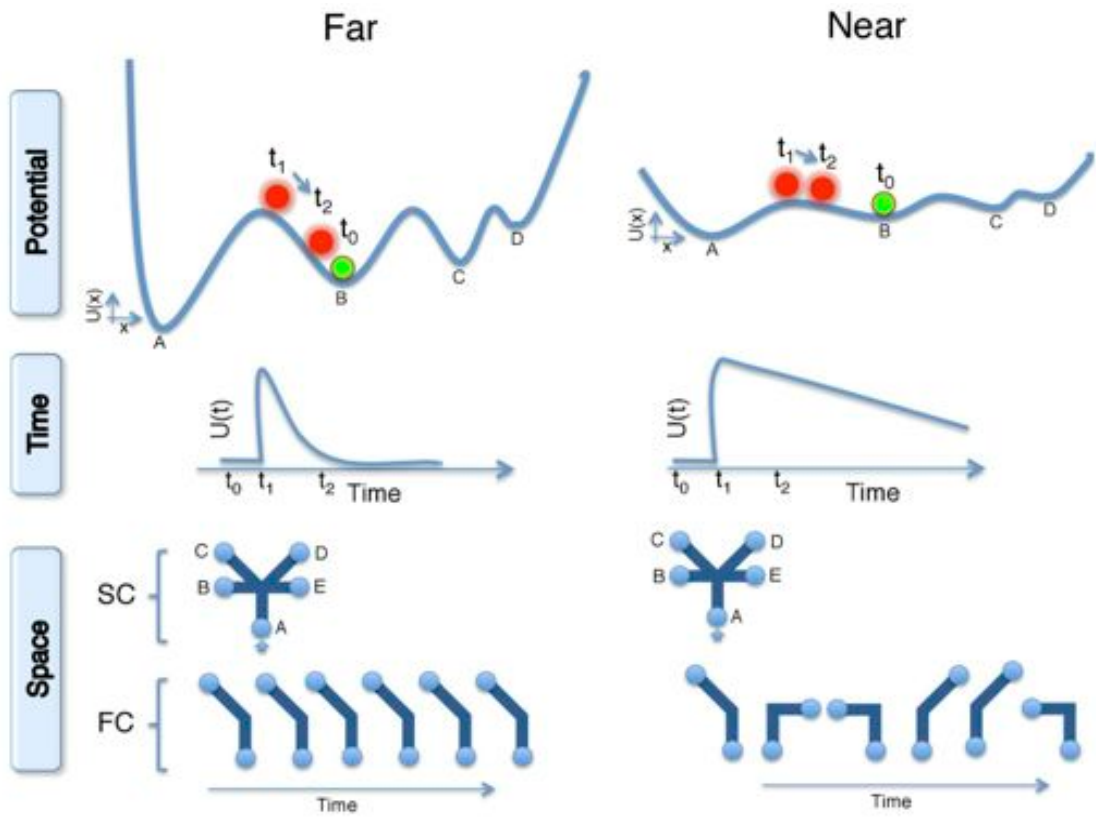